\documentclass[aps, twocolumn, showpacs, nofootinbib,superscriptaddress]{revtex4-1}
\usepackage{graphicx}
\usepackage{amsmath}
\usepackage{amsfonts}
\usepackage{amssymb}
\usepackage{amsthm}

\usepackage{hyperref}

\newcommand{\ket}[1]{|{#1}\rangle}
\newcommand{\bra}[1]{\langle{#1}|}

\newcommand{\ketbra}[2]{|{#1}\rangle\langle{#2}|}

\newcommand{\beq}{\begin{equation}}
\newcommand{\eeq}{\end{equation}}

\begin{document}

\author{Christian Arenz}
\affiliation{Theoretische Physik, Universit\"at des Saarlandes, D 66123 Saarbr\"ucken, Germany}

\author{Cecilia Cormick}
\affiliation{Theoretische Physik, Universit\"at des Saarlandes, D 66123 Saarbr\"ucken, Germany}
\affiliation{Institute for Theoretical Physics, Universit\"at Ulm, D 89081 Ulm, Germany}

\author{David Vitali}
\affiliation{School of Science and Technology, Physics Division, University of Camerino, Camerino (MC), Italy}

\author{Giovanna Morigi}
\affiliation{Theoretische Physik, Universit\"at des Saarlandes, D 66123 Saarbr\"ucken, Germany}

\title{Generation of two-mode entangled states by quantum reservoir engineering}

\date{\today}

\begin{abstract}
A method for generating entangled cat states of two modes of a microwave cavity field is proposed.
Entanglement results from the interaction of the field with a beam of atoms crossing the microwave resonator, giving
rise to non-unitary dynamics of which the target entangled state is a fixed point.  We analyse the robustness of the
generated two-mode photonic ``cat state''  against dephasing and losses by means of numerical simulation. This proposal
is an instance of quantum reservoir engineering of photonic systems.
\end{abstract}


\maketitle

\section{Introduction}

Quantum reservoir engineering generally labels a strategy at the basis of protocols which make use of the non-unitary
evolution of a system in order to generate robust quantum coherent states and dynamics \cite{Diehl_etal_NPhys_2008}. The
idea is in some respect challenging the naive expectation, that in order to obtain quantum coherent dynamics one shall
warrant that the evolution is unitary at all stages. Due to the stochastic nature of the processes which generate the
target dynamics, strategies based on quantum reservoir engineering are in general more robust against variations of the
parameters than protocols solely based on unitary evolution \cite{Diehl_etal_NPhys_2008,
Verstraete_etal_NPhys_2009,Kraus_etal_PRA_2008}. A prominent example of quantum reservoir engineering is laser cooling,
achieving preparation of atoms and molecules at ultralow temperatures by means of an optical excitation followed by
radiative decay \cite{Ya.B.Zeldovich_Wineland}.  The concept of quantum reservoir engineering and 
its application for quantum information processing has been formulated in Refs.  \cite{Cirac_PRL_1993,
Poyatos_etal_PRL_1996}, and further pursued in Refs. \cite{Plenio-PRA-1999, Carvalho_PRL_2001, Plenio-Huelga-PRL-2002}.
Proposals for quantum reservoir engineering of quantum states in cavity quantum electrodynamics
\cite{Pielawa_etal_PRL_2007,Susanne_et.al,Sarlette-PRL-2011,Kastoryano-PRL-2011,Jaksch-2012,
Everitt-2012} and many-body systems \cite{Diehl_etal_NPhys_2008,Verstraete_etal_NPhys_2009, Fogarty,Zippilli:2013}
have been recently discussed in the literature and first experimental realizations have been reported
\cite{Krauter_PRL2011,Rempe_Sience2008,Barreiro}. Applications for quantum technologies are being pursued
\cite{Vollbrecht_PRL2011,Barreiro,Goldstein,Huelga}.

\begin{figure}[htb]
 \includegraphics[width=170pt]{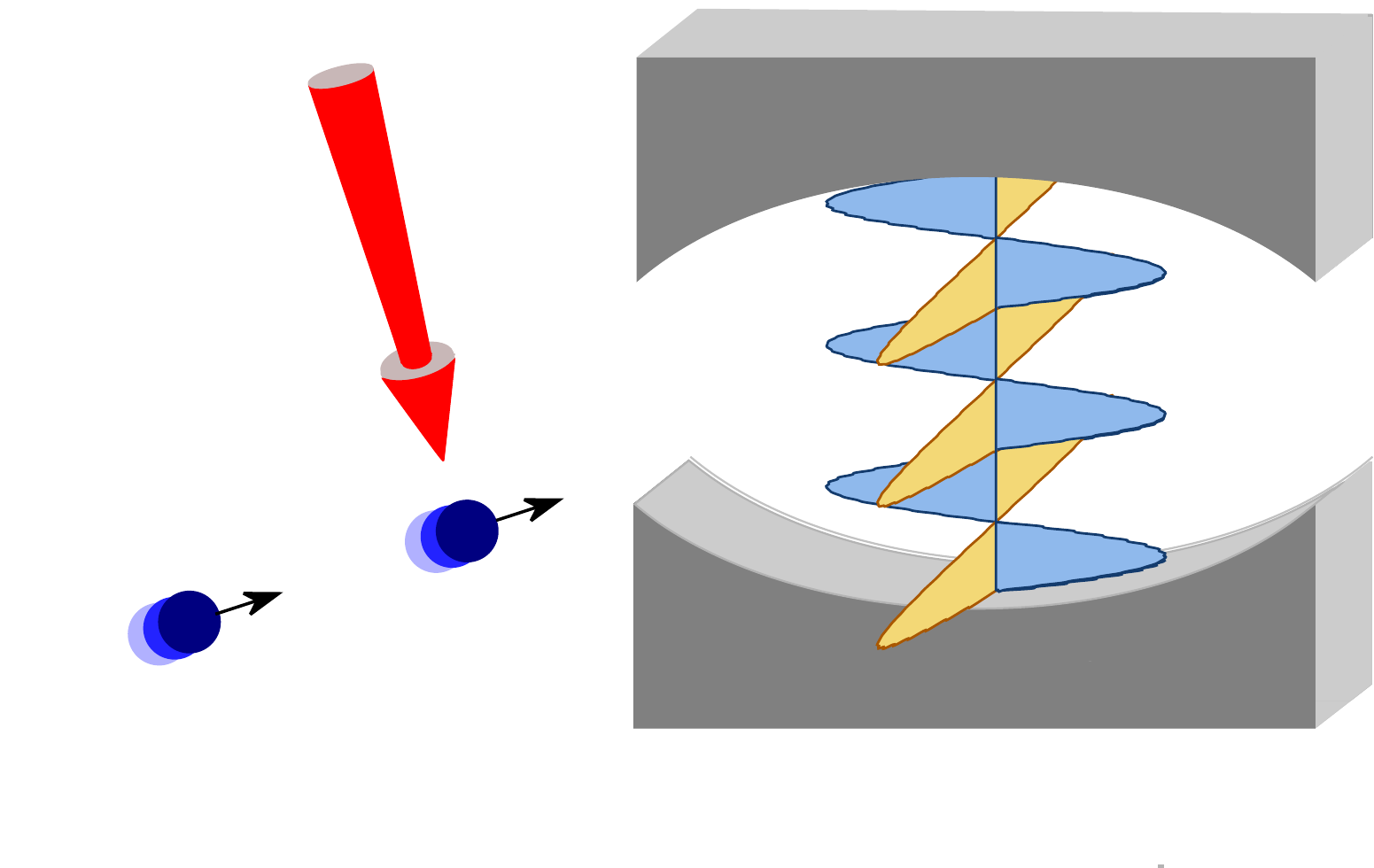}
 \caption{\label{fig:system} A high-finesse microwave resonator is pumped by a beam of atoms with random arrival times.
Two modes of the cavity are coupled to two atomic transitions, which are driven by external lasers while
interacting with the fields. The fields undergo non-unitary dynamics, whose asymptotic state is an entangled state as
in Eq.~\eqref{eq:target state}. These dynamics could be implemented in the experimental setup of
Ref.~\cite{Deleglise_Nat_2008}.}
\end{figure}

In this article we propose a protocol based on quantum reservoir engineering for preparing a cavity in a highly
nonclassical entangled ``cat-like'' state. This protocol is applicable to the experimental setup realized in
\cite{Raimond_RMP2001, Walther_RPP2006}, which is pumped by a
beam of atoms with random arrival times. In this setup the system dynamics intrinsically stochastic due to the
impossibility of controlling the arrival times of the atoms, but only their rate of injection, and the finite detection efficiency. The protocol we discuss allows one to generate and stabilize an entangled state of two modes of a microwave resonator,
by means of an effective environment constituted by the atoms. We show that when the internal state of the atoms
entering the cavity is suitably prepared and external classical fields couple the atomic transitions, then the
asymptotic state of the cavity modes takes the form
\beq
|\psi_\infty\rangle=(|\alpha\rangle_{\rm A}|\alpha\rangle_{\rm B} +|-\alpha\rangle_{\rm A}|-\alpha\rangle_{\rm
B})/\mathcal
N\,,
\label{eq:target state}
\eeq
where $|\alpha\rangle_j$ denotes a coherent state of mode $j=A,B$ with complex amplitude $\alpha$ and $\mathcal
N=\sqrt{2[1+\exp(-4|\alpha|^{2})]}$ is the normalization constant.

Our proposal extends previous works of some of us, which are focussed on generating two-mode squeezing in a microwave
cavity \cite{Pielawa_etal_PRL_2007} and entangling two distant cavities using a beam of atoms \cite{Susanne_et.al}.
The state of Eq.~(\ref{eq:target state}) whose robust generation is proposed here is not simply entangled but
possesses strongly nonclassical features, being a nonlocal macroscopic superposition state similar to those discussed
in Ref.~\cite{Davidovich1993}. The setup we consider is sketched in Fig. \ref{fig:system}, and is
similar to the one realized in Ref. \cite{Walther_RPP2006,Deleglise_Nat_2008}.

This work is structured as follows. In Sec. \ref{sec:proposal} we sketch the general features of our proposal.
Section \ref{sec:Lindblad} presents a method to engineer each of the target dynamics starting from the Hamiltonian of an
atom of the beam, which interacts with the cavity for a finite time. Results from numerical simulations are reported and
discussed in Sec. \ref{sec:results}. The conclusions are drawn in Sec. \ref{Sec:V}.

\section{Target master equation and asymptotic state} \label{sec:proposal}

Let $\rho$ be the density matrix for the degrees of freedom of the two cavity modes and
$\rho_\infty=|\psi_\infty\rangle\langle \psi_\infty|$ the target state we want to generate with $|\psi_\infty\rangle$ in
Eq. \eqref{eq:target state}. The purpose of this section is to derive the master equation
\begin{equation}
\frac{\partial}{\partial t}\rho=\mathcal L\rho\,,
\end{equation}
for which $\rho_\infty$ is a fixed point, namely,
\beq
\label{eq:L}
\mathcal L\rho_\infty=0\,.\eeq
In order to determine the form of the Lindbladian ${\mathcal L}$ we first introduce the operators $a$ and $b$ which
annihilate a photon of the cavity mode A and B, respectively. It is simple to show that $\rho_\infty$ is a
simultaneous right eigenoperator at eigenvalue zero of the Liouvillians
\beq \label{eq:Liouville operators}
\mathcal L_j\rho=\gamma_{j}(2C_{j}\rho C_{j}^{\dagger}-\{C_{j}^{\dagger}C_{j},\rho\}), \quad j=1,2
\eeq
with $\gamma_j$ rates which are model-dependent and where the operators $C_{j}$ read
\beq \label{eq: Lindblad operator}
C_{1}=\frac{a-b}{\sqrt{2}}, \quad
C_{2}=2(ab-\alpha^{2}).
\eeq
In fact, $|\psi_j\rangle$ is eigenstate of $C_1$ and $C_2$ with eigenvalue 0, $C_j|\psi_{\infty}\rangle=0$.
The procedure we will follow aims at constructing effective dynamics described by the Liouvillian
\beq
\label{L12}
\mathcal L=\mathcal L_1+\mathcal L_2
\eeq
by making use of the interaction with a beam of atoms.

Before we start, we shall remark on two important points. In first place, the state $\rho_{\infty}$ is not the unique
solution of Eq. \eqref{eq:L} when $\mathcal L=\mathcal L_1+\mathcal L_2$. Indeed, states $|\alpha\rangle_{\rm
A}|\alpha\rangle_{\rm B}$ and $|-\alpha\rangle_{\rm A}|-\alpha\rangle_{\rm B}$, and any superposition of these two
states, are also eigenstates of both $C_1$ and $C_2$ at eigenvalue zero. We
denote the corresponding eigenspace by $\mathcal H_d$, which is a subspace of the Hilbert space of all states of the two cavity modes. The most general stationary state of $\mathcal L$ can be written as a statistical mixture, $\rho^{ss}= \sum_d p_d \ketbra{\psi_d}{\psi_d}$  \cite{Footnote}, where the sum spans over all the states $\ket{\psi_d} \in \mathcal H_d$, and $p_d$ are real and positive scalars such that $\sum_d p_d=1$.

Nevertheless, for the evolution determined by the Lindbladian of Eq. \eqref{L12} the state $\rho_\infty$ is the unique
asymptotic state provided that the initial state is the vacuum state for both cavity modes,
$\rho_0=\ketbra{0_A,0_B}{0_A,0_B}$. This can be shown using the parity
operator defined as 
\beq
\Pi_+=(-1)^{c^{\dagger}_+c_+}
\eeq
with $c_\pm = (a\pm b)/\sqrt{2}$. Operator $\Pi_+$ commutes with the operators $C_1$ and $C_2$, since
\beq
\label{c:pm}
C_1 = c_-\,, \quad
C_2 = c_+^2-c_-^2-2\alpha^2\,.
\eeq
Therefore, if the initial state can be written as statistical mixture of eigenstates of $\Pi_+$ with eigenvalue $+1$,
the time-evolved state will also be a statistical mixture of eigenstates with eigenvalue $+1$, and so will be the steady
state. In particular, $|\psi_\infty\rangle$ is the only state of subspace $\mathcal H_d$ which is eigenstate of $\Pi_+$
with eigenvalue $+1$, namely, $\Pi_+|\psi_\infty\rangle=|\psi_\infty\rangle$, and thus, under this condition, the
asymptotic state will be pure and given by $\rho_{\infty}$. Here we will assume just this situation, i.e., that
the cavity modes are initially prepared in the vacuum state, which is an even eigenvalue of operator $\Pi_+$, and which
represents a very natural initial condition.

These considerations are so far applied to the ideal case in which the dynamics of the cavity modes density matrix are
solely determined by Liouvillian ${\mathcal L}$ in Eq. \eqref{L12}. In this article we will construct the dynamics in
Eq. \eqref{L12} using a beam of atoms crossing with the resonator, as it is usual in microwave cavity quantum
electrodynamics. We will then analyze the efficiency of generating state $\rho_{\infty}$ at the asymptotics of
the interaction of the cavity with the beam of atoms, taking also into account experimental limitations.

\section{Engineering dissipative processes} \label{sec:Lindblad}

Our starting point is the Hamiltonian for the coherent dynamics of an atom whose selected Rydberg transitions
quasi-resonantly couple with the cavity modes. The atoms form a beam with statistical Poissonian distribution in the
arrival times. The mean velocity determines the average interaction time $\tau$ during which each
atom interacts with the cavity field, while the arrival rate $r$ is such to warrant that $r\tau\ll 1$, namely, the
probability that two atoms interact simultaneously with the cavity is strongly suppressed. The master equation for the
density matrix $\chi$ describing the dynamics of the cavity modes coupled with one atom reads
\beq
\label{master:0}
\frac{\partial}{\partial t}\chi=\frac{1}{{\rm i}\hbar}[H,\chi]+\kappa{\mathcal K}\chi\,,
\eeq
with $H$ the Hamiltonian governing the coherent dynamics and
\begin{eqnarray}
\mathcal K \chi=2 a \chi a^\dagger + 2 b \chi b^\dagger - \{a^\dagger a, \chi\}- \{b^\dagger b, \chi\}
\end{eqnarray}
the superoperator describing decay of the cavity modes at rate $\kappa$. The field density matrix is found after tracing
out  the atomic degrees of freedom, and formally reads $\rho(t)={\rm Tr}_{\rm at}\{\chi(t)\}$.
In the following we will specify the form of Hamiltonian $H$ and derive an effective master equation for the density
matrix $\rho$ of the cavity field interacting with a beam of atoms, which approximates the dynamics governed the
Liouvillian $\mathcal L$ in Eq. \eqref{L12}.

In the following we shall analyze separately each of the processes corresponding to the two types of Lindblad
superoperators composing the sum in Eq.  \eqref{L12}. Note that cavity losses are detrimental, as they do not
preserve the parity $\Pi_+$ of the state of the cavity. In the rest of this section they will be
neglected, their effect will be considered when calculating numerically the efficiency of the protocol.

\subsection{Realization of the Lindblad superoperator ${\mathcal L}_1$.}

\begin{figure}[!h]
\centering
\includegraphics[width=165pt]{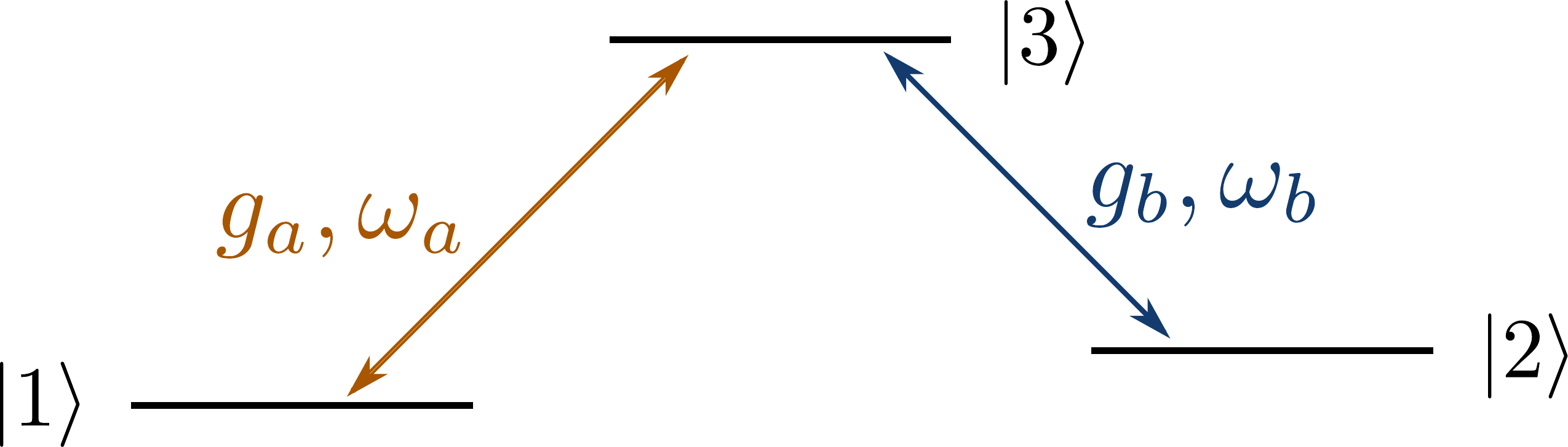}
\caption{Relevant atomic levels and couplings leading to the dynamics which realizes the Lindblad superoperator
$\mathcal L_1$. The atom is prepared in state $\ket{-}$, Eq. \eqref{eq:basis}.
\label{fig:L1}}
\end{figure}

We now show how to implement the dynamics described by the Lindblad superoperator $\mathcal L_1$. For this purpose, we
assume that the atomic transitions effectively coupling with the cavity modes form a  $\Lambda$-type configuration of
levels,  as schematically represented in Fig. \ref{fig:L1}.  The interaction of a single atom with the cavity modes is
governed by the Hamiltonian
\begin{eqnarray}
H& =&  \hbar\omega_a a^\dagger a +  \hbar\omega_b b^\dagger b +  \hbar\omega_2 \sigma_{2,2} + \hbar \omega_3
\sigma_{3,3} \\
& &+  \hbar(  g_a a^{\dagger}\sigma_{1,3} +   g_b b^{\dagger}\sigma_{2,3} +{\rm H.c.})\,,\nonumber
\end{eqnarray}
where $\omega_a$ and $\omega_b$ are the frequencies of the cavity modes, $\omega_2$ ($\omega_3$) is the energy of
level $\ket{2}$ ($\ket{3}$), here setting the energy of level $\ket{1}$ to zero, $g_a$ and $g_b$ are the vacuum Rabi
frequencies characterizing the strength of the coupling of the dipolar transitions $\ket{1}\to\ket{3}$ and
$\ket{2}\to\ket{3}$, respectively, with the corresponding cavity mode, and $\sigma_{j,k}=\ketbra{j}{k}$ is the spin-flip
operator. In the following we assume that the transitions are resonant, i.e. $\omega_a= \omega_3$ and $\omega_b =
\omega_3-\omega_2$.

In the reference frame rotating with the cavity modes, the Hamiltonian can be rewritten as
\begin{eqnarray}
H_1
%&=& \hbar  (g_a a^{\dagger}\sigma_{1,3} +g_b  b^{\dagger}\sigma_{2,3}+{\rm H.c.})\nonumber\\
= \hbar \sqrt{2} \frac{g_ag_b}{g}(c_-^{\dagger}\sigma_{-,3} +c_+^{\prime\dagger} \sigma_{+,3}+{\rm H.c.})\,,
\label{H_1}
\end{eqnarray}
where $g=\sqrt{g_a^2+g_b^2}$, $c_-$ is defined in Eq. \eqref{c:pm} and $\sigma_{\pm,3}=\ketbra{\pm}{3}$, with
\begin{align}
|-\rangle=\frac{g_b|1\rangle-g_a |2\rangle}{g}\,,\quad |+\rangle=\frac{g_a|1\rangle+g_b |2\rangle}{g}\,,
\label{eq:basis}
\end{align}
while $c_+'$ is a superposition of modes $a$ and $b$.  This representation clearly shows that, if the atoms are injected
in the state $\ket{-}$ and interact with the resonator for a time $\tau_1$ such that $g\tau_1\ll 1$, they may only
absorb photons of the ``odd'' mode $c_-$. More precisely, the condition to be fulfilled is $g\tau_1\sqrt{N_-+1/2}\ll1$,
where $N_-$ is the mean number of photons in the odd mode, $N_-=\langle c_-^\dagger c_-\rangle$. In this case, if
$\rho(t)$ is the state of the field at the instant in which an atom in state $\ket{-}$ is injected, the state of the
field $\rho$ at time $t+\tau_1$ reads  \beq
\rho(t+\tau_1)=\rho(t) + \frac{g_a^2g_b^2}{g^2}\tau_1^{2}
\left[2c_-\rho(t)c_-^{\dagger}-\{c_-^{\dagger}c_-,\rho(t)\}\right]\,.
\label{eq:master1}
\eeq
This corresponds to the desired process, which drives the odd mode into the vacuum state. Here, we neglect corrections
that are smaller by a factor of order $g^2\tau_1^2 (N_- +1/2)$.

Assuming that the atoms in state $\ket{-}$ are injected at rate $r_1$ with $r_1\tau_1\ll 1$, the probability of having
two atoms simultaneously inside the cavity can be neglected. In this case the field evolution can be analysed on a
coarsed-grained time scale $\Delta t$ such that $\Delta t \gg \tau_1$ and $r_1\Delta t\ll 1$. After expressing the
differential quotient $[\rho(t+\Delta t)-\rho(t)]/\Delta t$ as a derivative with respect to time one recovers the master
equation \cite{Susanne_et.al}
\begin{align}
\label{Master:1}
\frac{\partial}{\partial t}\rho(t)\simeq\gamma_1 \left[ 2c_-\rho(t)c_-^{\dagger}-\{c_-^{\dagger}c_-,\rho_(t)\} \right],
\end{align}
which corresponds to the dynamics governed by superoperator $\mathcal L_1$ in Eq. \eqref{eq:Liouville operators}. Here,
\beq
\gamma_1=r_1\frac{g_a^2g_b^2}{g^2}\tau_1^{2}\,.
\eeq

We note that Eq. \eqref{Master:1} is valid as long as higher order corrections are negligible. This condition provides
an upper bound to the rate $\gamma_1$, i.e., $\gamma_1\ll r_1$. However, it is not strictly necessary that the dynamics
take place in this specific limit: One can indeed speed up the process of photon absorption from the odd mode taking
longer interaction times between the atom and the cavity. In this case, the form of the master equation is different,
but one could obtain absorption of photons from the odd mode. We refer the reader to Ref. \cite{Susanne_et.al}, where
the required time has been characterized for a similar proposal in the different regimes.

\subsection{Realization of the Lindblad superoperator ${\mathcal L}_2$.} \label{subsec:Lindblad2}

\begin{figure}[!h]
\centering
\includegraphics[width=80pt]{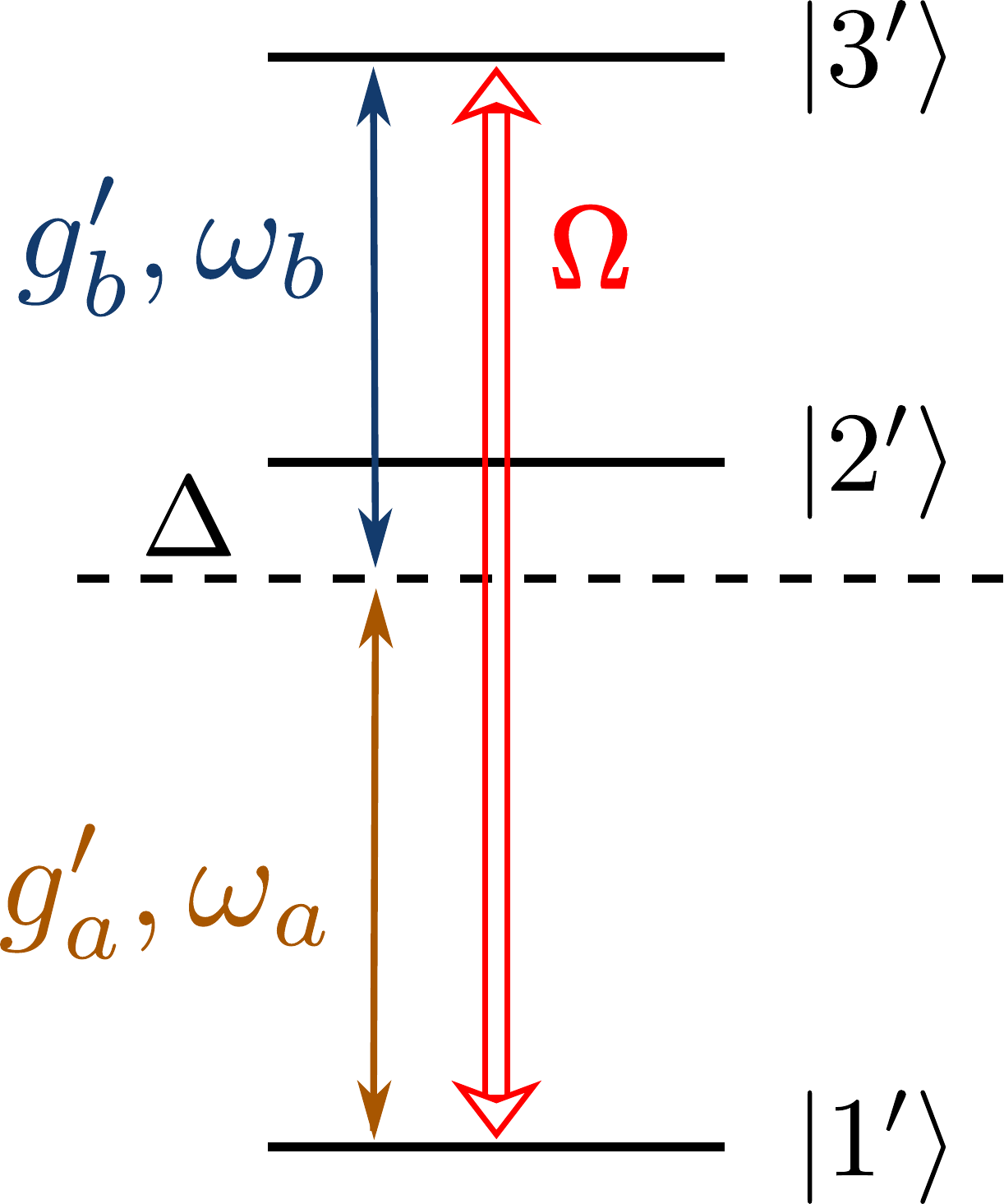}
\caption{Relevant atomic levels and couplings leading to the dynamics which approximates the Lindblad superoperator
$\mathcal L_2$. A classical field of amplitude $\Omega$ drives resonantly the transition $\ket{1'}\to\ket{3'}$. This
transition is also resonantly driven by two-photon processes, in which a photon of cavity mode A and a photon of cavity
mode B are simultaneously absorbed or emitted. These dynamics dominate over one-photon processes by choosing the
detuning $|\Delta|$ sufficiently larger than the coupling strengths $g_a',g_b'$.
\label{fig:L2}}
\end{figure}

The dynamics described by the Lindblad operator $\mathcal L_2$, Eq. \eqref{L12}, can be realized using a level scheme as
shown in Fig. \ref{fig:L2}.  We denote by $\omega_j'$ the frequency of the atomic state $\ket{j=2,3}$, such that
$\omega_3'>\omega_2'>\omega_1'=0$. The transition is such that $\omega_3'=\omega_a+\omega_b$.

A laser drives resonantly the transition $\ket{1'}\to\ket{3'}$, so that the frequency
$\omega_L=\omega_3'=\omega_a+\omega_b$. In the frame rotating at the
frequency of the cavity modes the Hamiltonian governing the coherent dynamics reads
\begin{multline}
\label{H_2}
 H_2=  \hbar  \Delta \sigma_{2'2'} +\hbar ( g_a' a^\dagger \sigma_{1'2'} + g'_b  b^\dagger \sigma_{2'3'}
 + \Omega \sigma_{1'3'} + {\rm H.c.}) \,,
\end{multline}
where $\Delta=\omega_2'-\omega_a$. We assume that $g_a'\sqrt{\langle n_a\rangle}, g_b'\sqrt{\langle n_a\rangle}\ll
|\Delta|$, with $\langle n_j\rangle$ the mean number of photons in the cavity mode $j=A,B$, and analyze the state of the
cavity field after it has interacted with an atom which is injected in state $\ket{1'}$. The interaction time is denoted
by $\tau$ and is chosen such that $|\Delta|\tau\gg1$ and $g_j'^2\langle n_j\rangle\tau/|\Delta|\ll 1$. The density
matrix for the cavity field at time $t+\tau$ can be cast in the form \cite{Susanne_et.al}
\begin{eqnarray}
\label{Master:L2}
\rho(t+\tau)=\rho(t)&+&\frac{1}{8} \left(\frac{g'_ag'_b\tau}{\Delta}\right)^2\Bigg[ 2C_2\rho
C_2^{\dagger}-\left\{C_2^{\dagger}C_2, \rho_f \right\}\Bigg]\nonumber\\
%This line was included by Christian%
&&+i\frac{g_{a}^{\prime 2}}{\Delta^2}\left(\Delta\tau - \sin\Delta \tau\right)[a^{\dagger}a,\rho]\nonumber\\
&&+2\frac{{g'_a}^2}{\Delta^2}\sin^2\left(\frac{\Delta \tau}{2}\right)\left(2a\rho a^\dagger -\{a^\dagger a,
\rho\}\right)\nonumber\\
&&-\frac{1}{2}\left(\frac{g_{a}^{\prime 2}\tau}{\Delta}\right)^2[a^{\dagger}a,[a^{\dagger}a,\rho]]                ,
\end{eqnarray}
where $\rho(t)$ is the density matrix before the interaction and $C_2=2(ab-\alpha^2)$, Eq. \eqref{eq: Lindblad
operator}. Here, $\alpha^2 = \Omega \Delta/ (g_a' g_b')$, showing that the number of photons at the asymptotics is
determined by $\Omega$. Equation \eqref{Master:L2} has been derived in perturbation theory and by tracing out the
degrees of freedom of the atom after the interaction. The first line of Eq. \eqref{Master:L2} describes two-photon
processes leading to the target dynamics at a rate determined by the frequency
$$\gamma_2^{(0)}=\frac{1}{8} \left(\frac{g_a^{\prime}g_b^{\prime}\tau}{\Delta}\right)^2\,,$$
while the terms in the other lines are unwanted processes, which occur at comparable rates and therefore lead to
significant deviations from the ideal behaviour. The second line of Eq. \eqref{Master:L2}, in particular, corresponds to
one-photon processes on the transition $\ket{1'}\to\ket{2'}$, leading to phase fluctuations of the cavity mode A. The
third line describes losses of mode A due to one-photon processes, and the last line gives dephasing effects of
cavity
mode A associated with two-photon processes. Other detrimental processes, leading to dephasing and amplification of the
field of cavity mode B, have been discarded under the assumption that the corresponding amplitude is of higher
order. This assumption is correct as long as the amplitude $\Omega$, determining the number of photons, is chosen to be
of the order of $g_j'^2/\Delta$ and fulfills the inequalities $(|\Delta|\tau)(\Omega\tau)\gg 1$ and $\Omega\tau\ll 1$.
This is therefore a restriction over the size of the cat state one can 
realize by means of this procedure.

Let us now discuss possible strategies in order to compensate the effect of the unwanted terms in Eq. \eqref{Master:L2}.
We first consider the term in the second line. This term scales with $g_a^{\prime 2}\tau/\Delta$ and is larger than
$\gamma_2^{(0)}$. It can be  compensated by means of a term of the same amplitude and opposite sign. This can be
realized by considering another atomic transition which is quasi resonant with the same cavity field, say, a third
transition $\ket{1_{\rm aux}}\to\ket{2_{\rm aux}}$ such that cavity mode A couples with strength $g_{\rm aux}$ and
detuning $\Delta_{\rm aux}$ with the dipolar transition with $|\Delta_{\rm aux}|\gg g_{\rm aux}$. If the atom is
prepared in the superposition $\cos(\varphi)\ket{1'}+\sin(\varphi)\ket{1_{\rm aux}}$ before being injected into the
cavity, then the coherent dynamics are governed by Hamiltonian $H_2'=H_2+h_{\rm aux}$, with
\beq
h_{\rm aux}=\hbar\Delta' \sigma_{2_{\rm aux}2_{\rm aux}} +\hbar g_{\rm aux} (a^\dagger \sigma_{1_{\rm aux}2_{\rm aux}} +
{\rm H.c.}) \,,
\eeq
which is reported apart for a global energy shift of the auxiliary levels.  It is thus  sufficient to select the
parameters so that the condition $\cos^2(\varphi)g_a^{\prime 2}/\Delta+\sin^2(\varphi)g_{\rm aux}^2/\Delta_{\rm aux}=0$
is
fulfilled, requiring that $\Delta$ and $\Delta_{\rm aux}$ have opposite signs.

This operation does cancel part of the dephasing due to the dynamical Stark shift of cavity mode A. It does not
compensate, however, the dephasing and dissipation terms due to one-photon processes and scaling with $g_a^{\prime
2}\/\Delta^2\sin\Delta\tau$ and $g_a^{\prime 2}/\Delta^2\sin^2(\Delta\tau/2)$, respectively. Nor does it cancel
the term due to two-photon processes in the last line of Eq. \eqref{Master:L2}, which scales with rate $(g_{a}^{\prime
2}\tau/\Delta)^2/2$. The remaining terms due to one-photon processes have a negligible effect for the choice of
parameters we perform, since $(g_a^{\prime 2}\/\Delta^2)/\gamma_2^{(0)}\sim (g_b'\tau)^{-2}$ and we choose $g_b'\tau\gg
1$ in order to warrant reasonably large rates (in other parameter regimes, where this is not fulfilled, these terms
could be set to zero by an appropriate selection of the velocity distribution of the injected atoms).

\begin{figure}[!h]
\centering
\includegraphics[width=130pt]{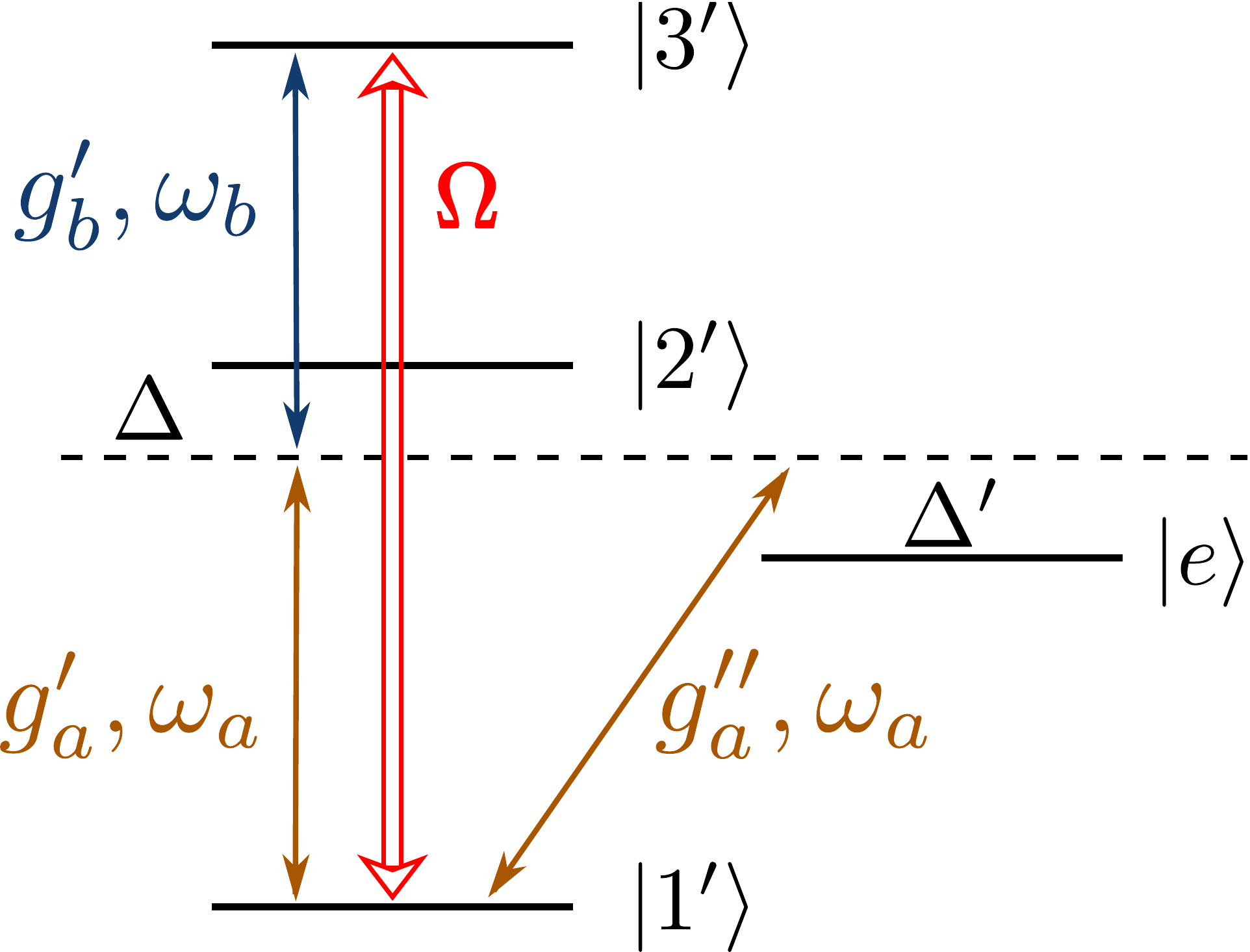}
\caption{Level scheme leading to the master equation \eqref{L2:ideal}. The coupling to the additional level $\ket{e}$
allows one to cancel out dephasing due to one-photon processes on transition
$\ket{1'}\to\ket{2'}$.
\label{fig:L2:1}}
\end{figure}

The last term can be made smaller than $\gamma_2^{(0)}$ when $(g_{b}^{\prime}/g_{a}^{\prime})^2\gg 1 $. Nevertheless,
this ratio cannot be increased arbitrarily, since the model we consider is valid as long as $\Omega\tau\ll 1$. This term
can be identically canceled out when specific configurations can be realized, like the one shown in Fig. \ref{fig:L2:1}:
In this configuration state $\ket{1'}$ couples simultaneously with the excited states $\ket{2'}$ and $\ket{e}$ by
absorption of a photon of mode A. The coherent dynamics are now described by Hamiltonian $H'=H_2+h'$ with
\beq
h'=\hbar\Delta' \sigma_{ee} +\hbar g_a''(a^\dagger \sigma_{1'e} + {\rm H.c.}) \,,
\eeq
If the coupling strengths and detunings are such that $g_a^{\prime 2}/\Delta=-g_a^{'' 2}/\Delta'$, then not only the
dynamical Stark shift cancels out, but interference in two-photon processes lead to the disappearance of the last line
in Eq. \eqref{Master:L2}. Under this condition, the resulting master equation is obtained in a coarse-grained time scale
$\Delta t$ assuming the atoms are injected in state $\ket{1'}$ at rate $r_2$ with a velocity distribution leading to a
normalized distribution $p(\tau)$ over the interaction times $\tau$, with mean value $\tau_2$ and variance $\delta\tau$
such that $\Delta t>\tau_2+\delta\tau$. For $r_2\Delta t\ll1$ the master equation reads
\begin{eqnarray}
\label{L2:ideal}
\frac{\partial}{\partial t}\rho&=&\gamma_2 \Bigg[ 2C_2\rho C_2^{\dagger}-\left\{C_2^{\dagger}C_2, \rho_f
\right\}\Bigg]\\
&&-{\rm i}f_1[a^\dagger a, \rho]
+f_2\left(2a\rho a^\dagger -\{a^\dagger a, \rho\}\right) , \nonumber
\end{eqnarray}
with $$\gamma_2 = (r_2/8) (g_a'g_b'/\Delta)^2 (\tau_2^2+\delta\tau^2)\,,$$ and
\begin{eqnarray}
&&f_1=r_2\frac{{g'_a}^2}{\Delta^2} \int_0^{\Delta t}{\rm d}\tau p(\tau)\sin(\Delta \tau)\,,\\
&&f_2=r_2\frac{{g'_a}^2}{\Delta^2} \int_0^{\Delta t}{\rm d}\tau p(\tau)\sin^2\left(\frac{\Delta \tau}{2}\right)\,.
\end{eqnarray}
When $p(\tau)$ is a Dirac-$\delta$ function, namely, $\delta\tau\to 0$, and $\tau_2\Delta=2n\pi$ with
$n\in\mathbb{N}$, then $f_1$ and $f_2 $ vanish identically and the 
dynamics describes the target Liouville operator. Under the condition that $\delta\tau\neq 0$, but $\epsilon\equiv
\Delta\delta\tau\ll 2\pi$, then $f_1={\rm O}(\epsilon^3)$ while $f_2=\epsilon^2/4$. In the other limit, in which
$p(\tau)$ is a flat distribution over $[0,2\pi/\Delta]$, then $f_1$ vanishes while $f_2\to 1/2$.

\subsection{Discussion}

In this section we have shown how to generate the target dynamics by identifying atomic transitions and initial states
for which the desired multiphoton processes are driven. The level schemes we consider could be the effective transitions tailored by means of
lasers. If the cavity modes to entangle have the same polarization  but different frequencies, the levels which are
coupled can be circular Rydberg states, while the coupling strengths $g_j$ can be effective transition amplitudes,
involving cavity and/or laser photons. The scheme then requires the ability to tune external fields so as to address
resonantly two or more levels, together with the ability to prepare the internal state of the atoms entering the
resonator. Depending on the initial atomic state, then, the dynamics can follow either the one described by
superoperator $\mathcal L_1$ or $\mathcal L_2$. An important condition is that no more than a 
single atom is present inside the resonator, which sets the bound over the total injection rate, $(r_1+r_2)\Delta t\ll
1$. The other important condition is that the dynamics are faster than the decay rate of the cavity. For the
experimental
parameters we choose, this imposes a limit, among others, on the choice of the ratio $g_j/|\Delta|$, determining both
the rate for reaching the ideal steady state as well as the mean number of photons per each mode, i.e., the size of the
cat.

\section{Results} \label{sec:results}

We now evaluate the efficiency of the scheme, implementing the dynamics given by Eq. \eqref{master:0} with $H=H_1+H_2'$,
where $H_1$ is given in Eq.  \eqref{H_1} and $H_2'=H_2+h$, with $H_2$ given in Eq. \eqref{H_2} while $h$ depends on the
additional levels which are included in the dynamics in order to optimize it. The initial state of the cavity is the
vacuum, and the atoms are injected with rate $r_1$ in state $\ket{1}$ (thus undergoing the coherent dynamics governed by
$H_1$) and with rate $r_2$ in state $\ket{\tilde{1}}$, which depending on the considered scheme can be either (i)
$\ket{1'}$ when $h=h'$, or (ii) $\cos(\varphi)\ket{1'}+\sin(\varphi)\ket{1''}$, when $h=h_{\rm aux}$. The case $h=0$ is
not reported,
since the corresponding efficiency is significantly smaller than the one achievable in the other two cases. In
order to determine the efficiency of the scheme we display the fidelity, namely, the overlap between the density matrix
$\chi(t)$ and the target state $\ket{\psi_\infty}$ as a function of 
the elapsed time. This is defined as
$$\mathcal F(t) = \bra{\psi_\infty} {\rm Tr}_{\rm at}\{\chi(t)\} \ket{\psi_\infty}\,,$$
where $\chi(t)$ is the density matrix of the whole system, composed by cavity modes and atoms of the beam which have
interacted with the cavity at time $t$, and ${\rm Tr}_{\rm at}$ denotes the trace over all atomic degrees of freedom.

\begin{figure}[h]
% (a)\phantom{\hspace{200pt}}
% \vspace{2pt}\\
 \includegraphics[width=0.85\columnwidth]{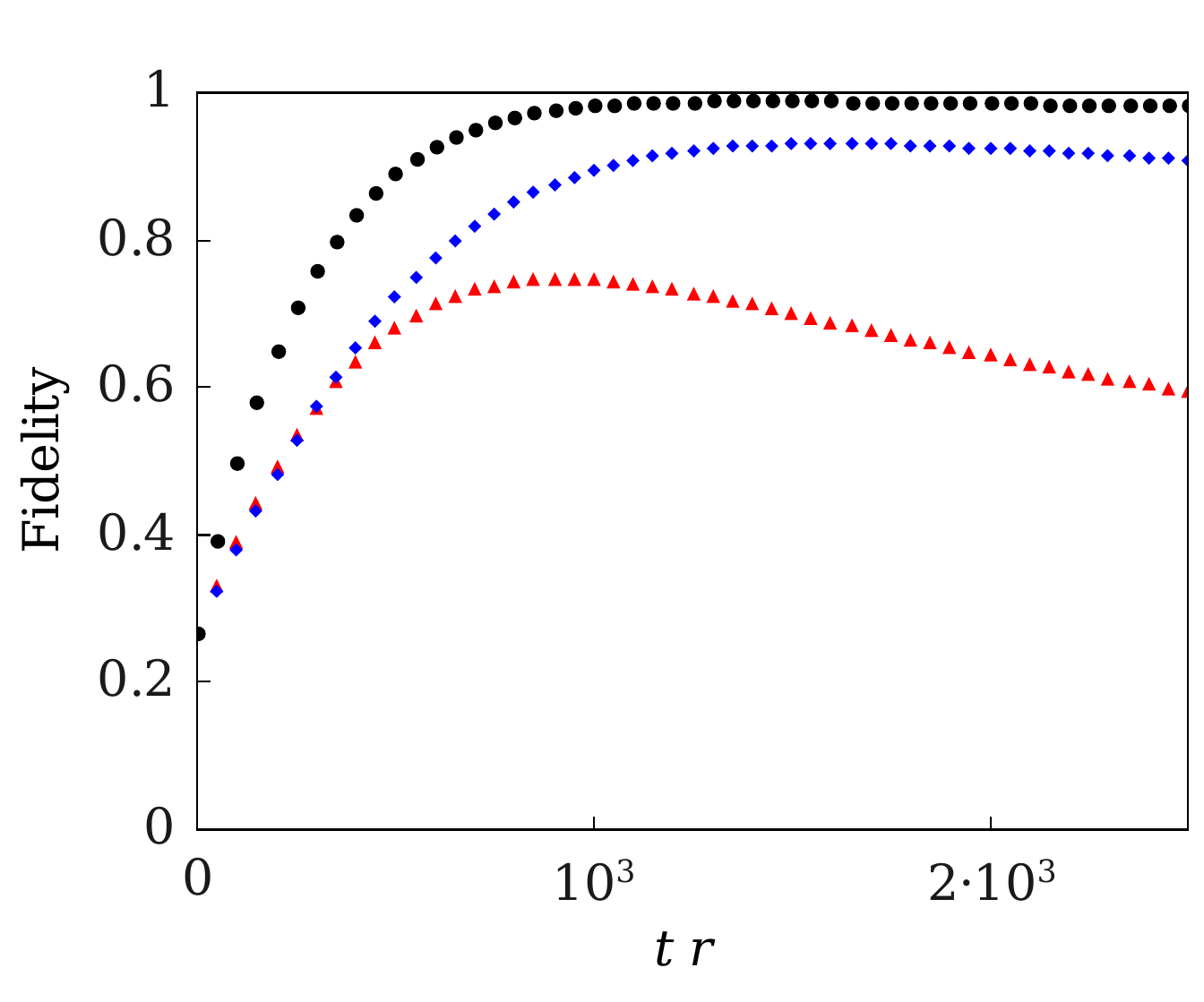}\\
% (b)\phantom{\hspace{200pt}}
%  \vspace{2pt}\\
% \includegraphics[width=0.85\columnwidth]{images/LogNeg_ideal.pdf}
 \caption{\label{fig:comparison} Fidelity as a function of time (in units of the injection rate $r=r_1=r_2$) for
$\alpha=1$, obtained by integrating numerically Eq. \eqref{master:0} after setting the cavity losses to zero,
$\kappa=0$. The other parameters are $g_{a}\tau_{1}=g_b\tau_1=0.1,~g_{b}'\tau_{2}=10^{2},~g_{b}'/\Delta=10^{-3}$,
$\Omega\tau_2=0.1$. From top to bottom: The black curve refers to $H_2'=H_2+h'$ with $g_a'=g_b'$, the other curves to $H_2'=H_2+h_{\rm aux}$
with $g_b'=10g_a'$ (blue) and $g_b'=3g_a'$ (red).
}
\end{figure}

For the purpose of identifying the best parameter regimes, we first analyze the dynamics neglecting the effect of cavity
losses. Figure \ref{fig:comparison} displays the fidelity as a function of time when the dynamics are governed by
Hamiltonian $H=H_1+H_2'$ for different realizations of $H_2'$ and for different parameter choices, when the amplitude of
the coherent state $\alpha=1$. Values of $\mathcal F\simeq 0.99$ are reached when $H_2'=H_2+h'$ is implemented. The
fidelity then slowly decays due to higher order effects, which become relevant at longer times. The effect of two-photon
processes involving mode A (which identically vanish for $H_2'=H_2+h'$) is visible in the two other curves, which
correspond to the dynamics governed by $H_2'=H_2+h_{\rm aux}$ when $g_b'=10g_a'$ (blue curve) and $g_b'=3g_a'$ (red
curve).  A comparison between these two curves shows that detrimental two-photon processes can be partially suppressed
by choosing the coupling rate $g_a'$ sufficiently smaller than $g_b'$.

\begin{figure}[h]
% (a)\phantom{\hspace{200pt}}
% \vspace{2pt}\\
 \includegraphics[width=0.85\columnwidth]{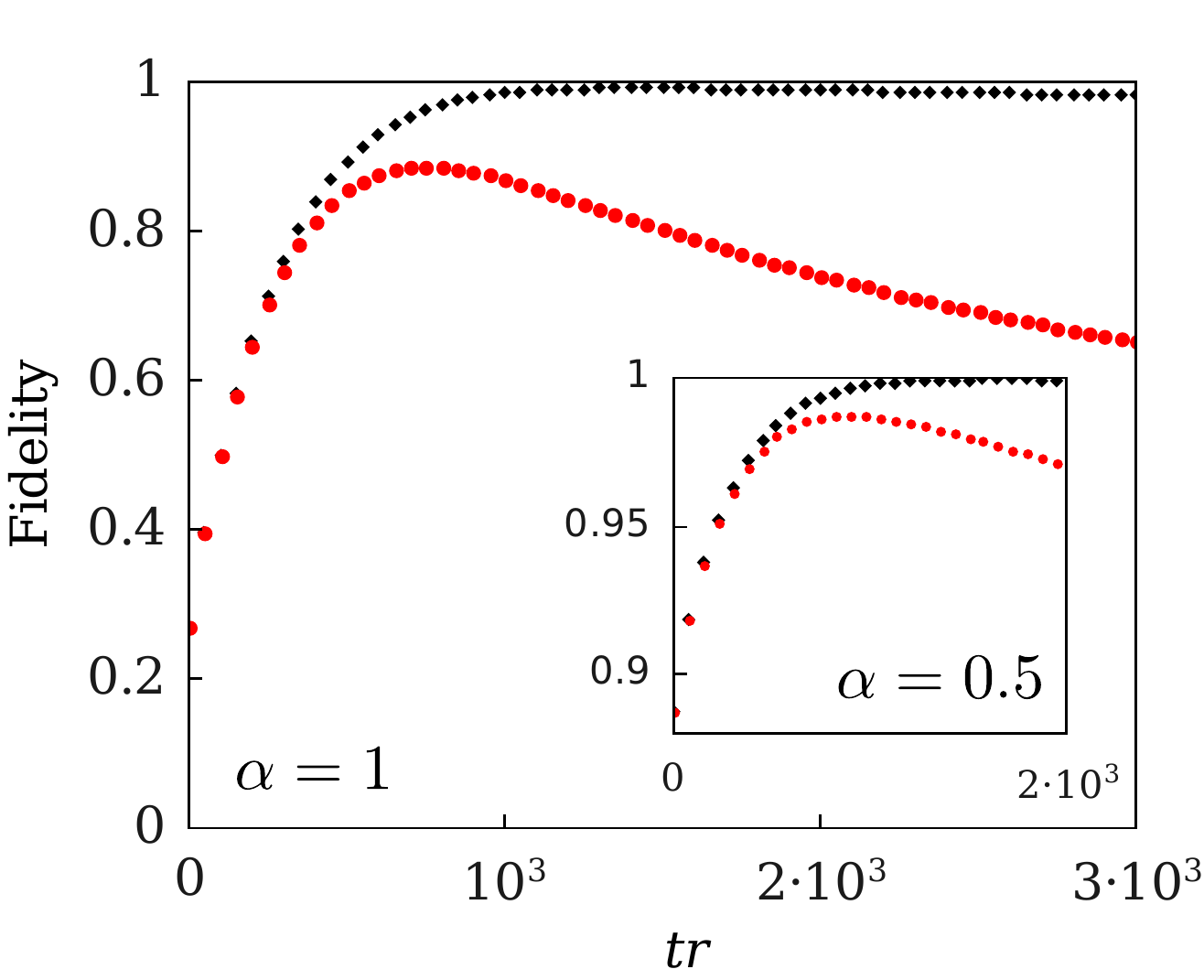}\\
% (b)\phantom{\hspace{200pt}}
%  \vspace{2pt}\\
% \includegraphics[width=0.85\columnwidth]{images/LogNeg_ideal.pdf}
 \caption{\label{fig:ideal} (a)  Fidelity as a function of time (in units of the injection rate $r=r_1=r_2$) for
$\alpha=1$, obtained by integrating numerically Eq. \eqref{master:0} after setting the cavity losses to zero,
$\kappa=0$. The other parameters are $\Omega\tau_2=0.1$, $g_{a}\tau_{1}=g_b\tau_1=0.1$,  whereby the black curve is
evaluated for $g_{b}'\tau_{2}=g_a'\tau_2=10^{2}$ and $g_{b}'/\Delta=10^{-3}$, while the red curve corresponds to
$g_{b}'\tau_{2}=g_a'\tau_2=10$ and $g_{b}'/\Delta=10^{-2}$ (from top to bottom). The inset has been evaluated for the
same parameters except for $\Omega\tau_2=0.05$, leading to $\alpha=0.5$.
}
\end{figure}

Figure \ref{fig:ideal} displays in detail the optimal case where $H_2'=H_2+h'$. The fidelity for the parameter choices
$g_{b}'/\Delta=10^{-3}$ and $g_{b}'/\Delta=10^{-2}$ are reported, showing that a smaller ratio leads to larger fidelity
in absence of cavity decay. The inset shows the corresponding fidelity when $\alpha=0.5$, which is notably
larger: Reaching this target state starting from the vacuum, in fact, requires a shorter time, for which higher-order
corrections are still irrelevant.

\begin{figure}[h]
 (a)\phantom{\hspace{200pt}}
 \vspace{2pt}\\
 \includegraphics[width=0.85\columnwidth]{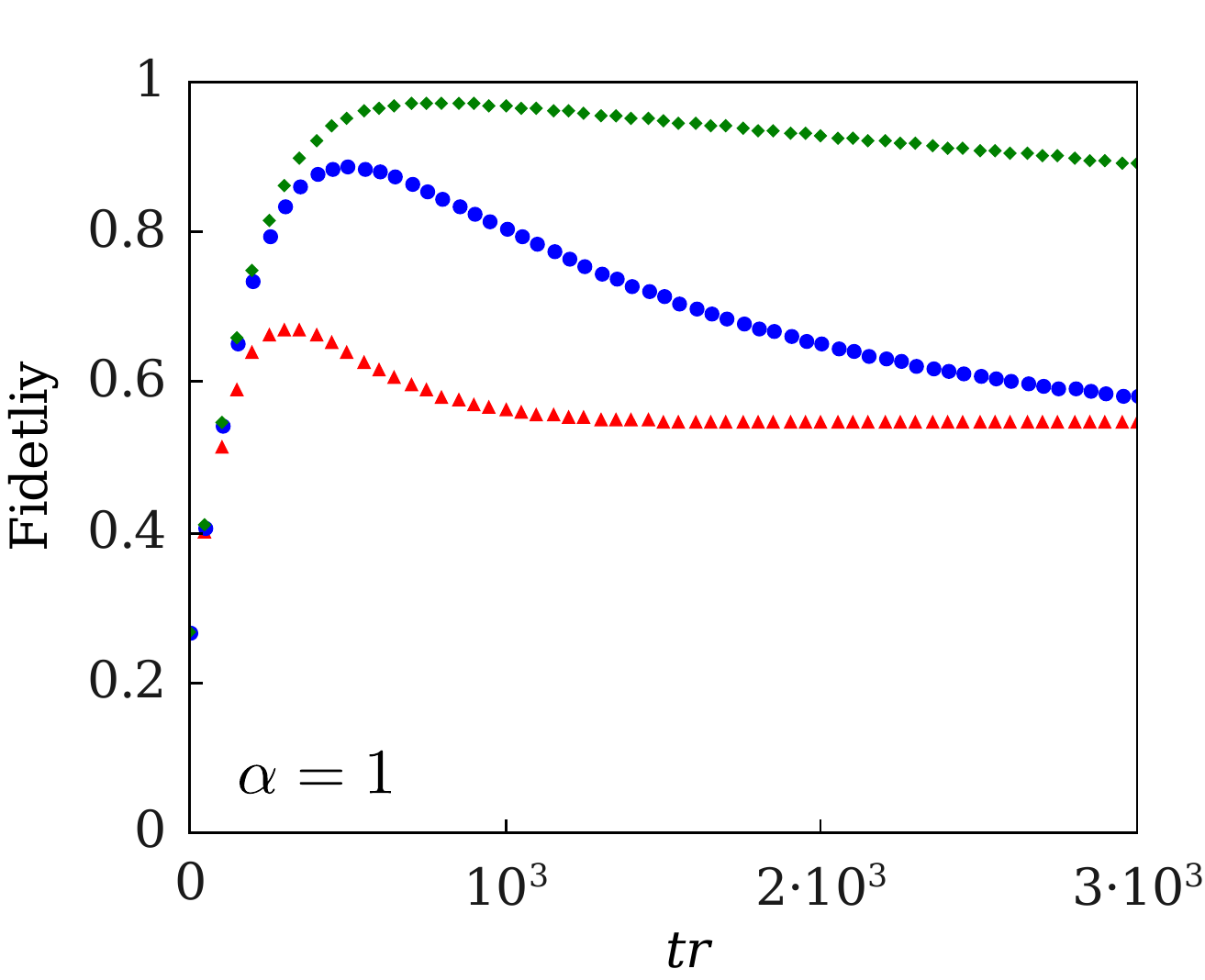}\\
 (b)\phantom{\hspace{200pt}}
 \vspace{2pt}\\
 \includegraphics[width=0.85\columnwidth]{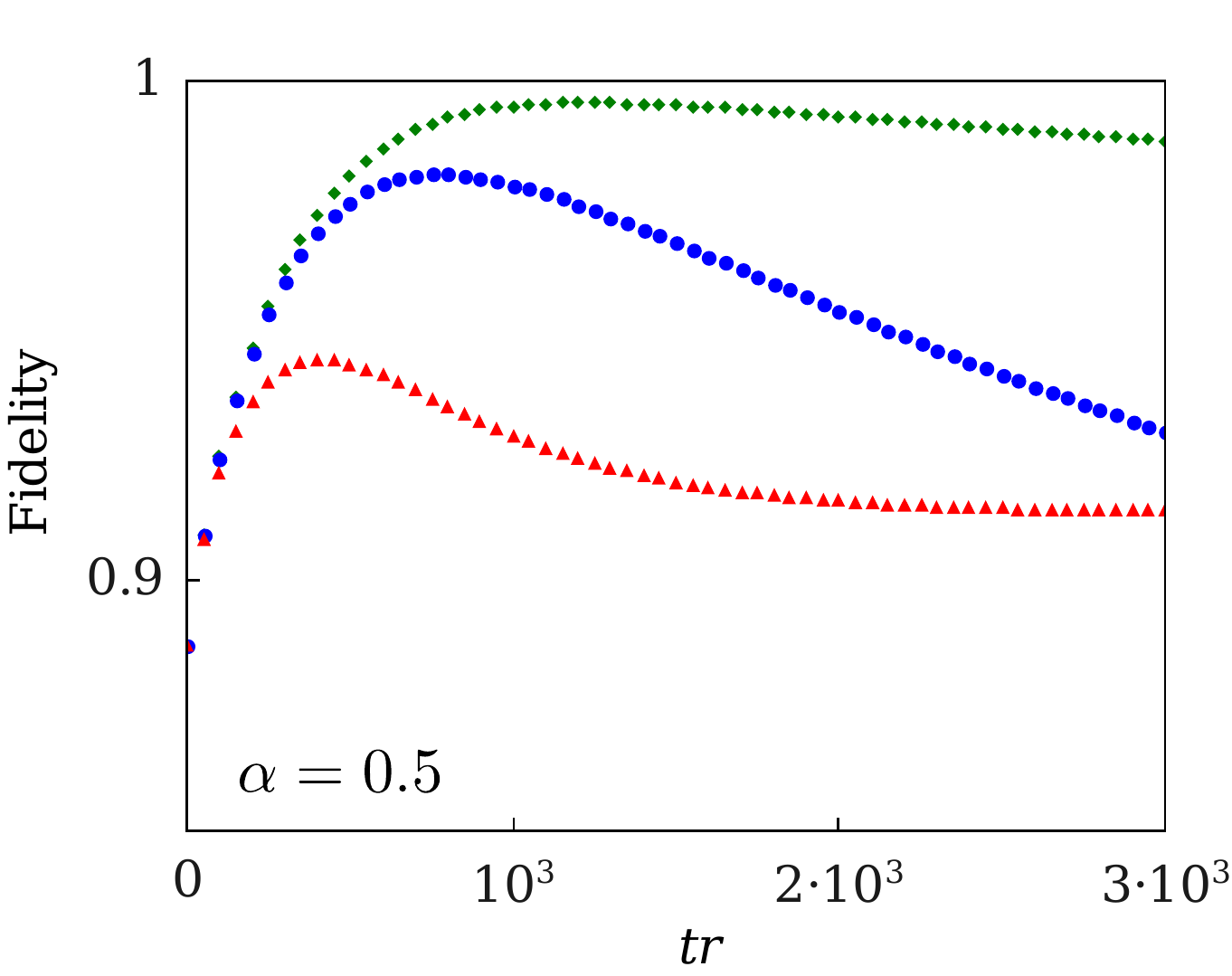}\\
 \caption{\label{fig:losses} Fidelity as a function of time for (a) $\alpha=1$ and (b) $\alpha=0.5$. The parameters are
the same as for the black curve in Fig. \ref{fig:ideal}, except that now cavity decay is included in the dynamics. In
particular, the green curve corresponds to $\kappa/r=10^{-5}$, the blue curve to~$\kappa/r=10^{-4}$, and the red curve
to $\kappa/r=10^{-3}$ (from top to bottom). The plots were obtained by integrating numerically Eq. \eqref{master:0}.}
\end{figure}

The effect of cavity losses is accounted for in Fig. \ref{fig:losses}, where the full dynamics of master equation
\eqref{master:0} is simulated  when $H_2'=H_2+h'$ and for different choices of the ratio $\kappa/r$. One clearly
observes
that the effect of cavity losses can be neglected over time scales of the order of $10^{-2}/\kappa$, so that
correspondingly larger rates $\gamma_1$ and $\gamma_2$ are required. Considered the parameter choice, this is possible
only by increasing the injection rate $r$. However, this comes at the price of increasing the probability that more than
one atom is simultaneously inside the resonator, thus giving rise to further sources of deviation from the ideal
dynamics.

These results show that degradation due to photon losses poses in general a problem to attain the target state
(\ref{eq:target state}): the rate of photon losses sets a maximum achievable fidelity, and also determines a time window
during which the fidelity is close to the maximum, after which the entanglement is gradually lost. The effect of the
photon losses is twofold: it leads to a decrease in the mean photon number, and also breaks the symmetry preservation in
the evolution. The decrease in the mean photon number can be compensated by increasing the strength $\Omega$ of the
pumping in the implementation of the second Lindblad operator, as long as the approximations made in Section
\ref{subsec:Lindblad2} are still valid.

\section{Concluding remarks}
\label{Sec:V}

A strategy has been discussed which implements non-unitary dynamics for preparing a cavity in an entangled state. It is
based on injecting a beam of atoms into a cavity, where the coherent interaction of the atoms with the cavity is a
multiphoton process pumping in phase photons, so that the cavity modes approach asymptotically the entangled state of
Eq. \eqref{eq:target state}. The procedure is robust against fluctuations of the number of atoms and interaction times.
It is however sensitive against cavity losses: the protocol is efficient, in fact, as long as the time scale needed in
order to realize the target state is faster than cavity decay. The effect of the photon losses is twofold: it damps the
mean photon number and also changes the parity of the state. It could be possible to partially revert the process by
measuring the parity of the total photon number and then performing a feedback mechanism, similar to the one proposed in
Refs.~\cite{Zippilli2003,Zippilli2004} and which has been partially 
implemented in Refs.~\cite{Sayrin2011,Zhou2012}. Alternatively, one can find a dissipative way to stabilize a unique
entangled target state without the need for feedback. This would require a process that can stabilize the parity of the
photon number in the even mode. First studies have been performed showing some increase in the final fidelity. We
finally note that these ideas could also find application in other systems, such as circuit quantum electrodynamics
setups \cite{E.Solano_PRA2002}.

\acknowledgements

We gratefully acknowledge discussions with Luiz Davidovich, Bruno Taketani, and Serge Haroche. This work was supported
by the European Commission (IP AQUTE, STREP PICC), by the BMBF QuORep, by the Alexander-von-Humboldt Foundation, and by
the German Research Foundation.


\begin{thebibliography}{99}
\bibitem{Footnote}
We do not have a proof that these states are the only stationary states of Eq. \eqref{eq:L}, where the Lindbladian is
given by Eq. \eqref{L12}. We have however checked numerically that they are the only asymptotic states for typical
experimental parameters, which is sufficient for the purposes of this paper.
\end{thebibliography}
\end{document}